\documentclass[twoside,a4paper,11pt]{sea10}
\usepackage{graphicx}
\usepackage{hyperref}
\begin{document}
\pagenumbering{arabic}
\pagestyle{myheadings}
\thispagestyle{empty}
{\flushleft\includegraphics[width=\textwidth,bb=58 650 590 680]{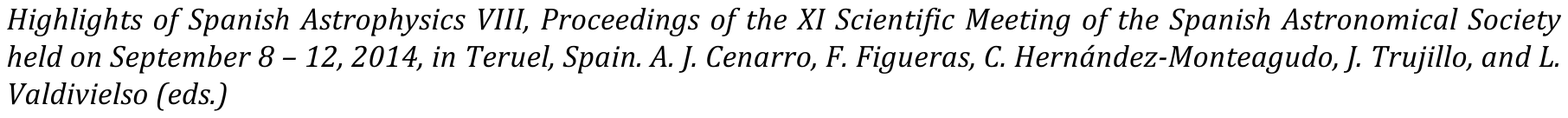}}
\vspace*{0.2cm}
\begin{flushleft}
{\bf {\LARGE
%
%%% TITLE of the paper. 
%%% TITLE of the paper. 
CAF\'E-BEANS: An exhaustive hunt for high-mass binaries
%
% Do not delete next few lines
}\\
\vspace*{1cm}
%
%%% Include here the LIST OF AUTHORS.
%%% Include here the LIST OF AUTHORS.
%%% Note that the last author has to be preceeded by an AND.
I. Negueruela$^{1}$, J. Ma\'{\i}z Apell\'aniz$^{2}$, S. Sim\'on-D\'{\i}az$^{3}$, E.~J. Alfaro$^{4}$, A. Herrero$^{3}$, J. Alonso$^{1}$, R. Barb\'a$^{5}$, J. Lorenzo$^{1}$, A. Marco$^{1}$, M.~Mongui\'o$^{1}$, N.~Morrell$^{6}$, A.~Pellerin$^{7}$, A.~Sota$^{5}$,
 and N.~R.~Walborn$^{8}$
%
% Do not delete next few lines
}\\
\vspace*{0.5cm}
%
%%% AFFILIATIONS LIST.
%%% and the AFFILIATIONS LIST. Note that one affiliation per line.
%%% Add as many affiliations as necessary. 
$^{1}$
DFISTS, Universidad de Alicante, Spain\\
$^{2}$
CAB-CSIC, Spain\\
$^{3}$
Instituto de Astrof\'{\i}sica de Canarias \& Universidad de La Laguna, Spain\\
$^{4}$
Instituto de Astrof\'{\i}sica de Andaluc\'{\i}a, CSIC, Spain\\
$^{5}$
Departamento de F\'{\i}sica, Universidad de La Serena, Chile\\
$^{6}$
Las Campanas Observatory, Chile\\
$^{7}$
SUNY Geneseo, USA\\
$^{8}$
Space Telescope Science Institute, USA
%
% Do not delete next few lines
\end{flushleft}
%
% Headings
\markboth{
%%% Type the SHORT version of the paper title.
%%% Type the SHORT version of the paper title.
CAF\'E-BEANS
}{ % Do not delete
%
%%%  First Author \& Second Author   OR   First-author et al. 
%%%  First Author \& Second Author   OR   First-author et al. if the author list 
%%% contains three or more authors.
I.~Negueruela et al.
% 
% Do not delete next few lines
}
\thispagestyle{empty}
\vspace*{0.4cm}
\begin{minipage}[l]{0.09\textwidth}
\ 
\end{minipage}
\begin{minipage}[r]{0.9\textwidth}
\vspace{1cm}
\section*{Abstract}{\small
%
% ABSTRACT ABSTRACT ABSTRACT
% ABSTRACT ABSTRACT ABSTRACT
%%% Type the ABSTRACT of your paper
CAF\'E-BEANS is an on-going survey running on the 2.2~m telescope at Calar Alto. For more than two years, CAF\'E-BEANS has been collecting high-resolution spectra of early-type stars with the aim  of detecting and characterising spectroscopic binaries. The main goal of this project is a thorough characterisation of multiplicity in high-mass stars by detecting all spectroscopic and visual binaries in a large sample of Galactic O-type stars, and solving their orbits. Our final objective is eliminating all biases in the high-mass-star IMF created by undetected binaries.
%
% Do not delete next few lines
\normalsize}
\end{minipage}
%
%
%%% BODY of the paper
%%% BODY of the paper
%
\section{Introduction \label{intro}}
High-mass stars play a crucial role in the dynamical and chemical
evolution of galaxies and their study is recognised as a keystone in modern astrophysics \cite{mas}. In spite of many important advances in recent years, we are still far from possessing firm knowledge of their evolutionary paths. Major uncertainties still remain in the
mass determination of O stars \cite{wv10}, and hence their radii, luminosities, and wind momenta \cite{stern}, parameters that determine the impact of massive stars on their environments and, through the number of ionising
photons, may affect even theories about re-ionisation in the
early Universe.

One of the main unknowns in our understanding is the role of binarity in high-mass star evolution \cite{hunter}. Multiplicity is known to be very high amongst high-mass stars, a fact that may be related to their formation mechanisms \cite{zy07}. Recent theoretical work has shown that evolution is very different for high-mass stars in binaries when compared to isolated ones \cite{eldstan,sana12}. The formation of most Wolf-Rayet stars may require binary-mediated mass-loss \cite{clark11}. Even the final fate of the star (the kind of supernova explosion it undergoes) depends on binarity \cite{claeys}.

Binarity studies have been carried out for some massive clusters \cite{ritchie09,kiminki}, finding very high fractions of binaries. But some other less dense clusters show binarity fractions not much higher than 20\% \cite{mahy09}. Recent estimates place the overall binary fraction at $\sim70$\% \cite{sana12}. The binary fraction amongst high-mass stars has a fundamental impact on the determination of the IMF. Star counts in resolved stellar populations are heavily affected by binarity, because of the steep luminosity function (e.g.\ a binary containing two 30\,$M_{\odot}$ stars will be counted as a single 40\,$M_{\odot}$ star). The effect is even stronger in unresolved populations, where high-mass binaries contribute to the radial velocity dispersion, used to estimate the IMF. An accurate determination of the binary fraction, the mass ratio ($q$) distribution in binaries and their period distribution is fundamental if we are to correct IMF determinations for these sources of uncertainty. Until we have a firm grasp on these numbers, the observed universality of the IMF has to be considered provisional, as top-heavy IMFs may be masked by the effects of binarity.

A second major issue under debate is the formation mechanism of high-mass stars. Though indirect evidence for the presence of accretion disks around some O-type stars exists (e.g.\ \cite{beltran}), it is still unclear if more massive stars form in the same way as low-mass stars, especially in view of their universal tendency to concentrate at the cores of very massive clusters. The statistics of multiplicity can help constrain formation mechanisms.

Most massive stars are believed to be born in multiple systems but many of them are still unidentified as such. Of the stars classified as spectroscopic binaries (SBs), only a small fraction have had their orbital period determined. Unfortunately, accurate stellar masses can only be obtained through the study of detached double-lined spectroscopic binaries (SB2) with favourable geometries (preferably eclipsing). The number of such systems among bright massive stars is frustratingly small. There is not much more than a dozen detached binaries with non-evolved components providing 
sufficiently stringent constraints on the masses of O-type stars
\cite{gies12}. 

In this context, several observational programmes have been started to cover the whole parameter space of high-mass binaries. Some of them try to detect visual binaries with small separations, by resorting to HST observations \cite{aldo14} or lucky imaging techniques, like those using AstraLux from Calar Alto and AstraLux Sur from La Silla (see \cite{maiz10}). For shorter orbital periods, when the components cannot be separated, spectroscopic observations to search for radial velocity shifts become the primary detection method. The most important survey for radial velocity variations among O-type stars is OWN, led by R.~Barb\'a \cite{barba10}. It has observed more than 200 massive stars (including 200 O stars) with declinations $<+20^{\circ}$. So far, it has discovered 126 new radial-velocity variables, of which 101 are confirmed binaries (53 new SB1, 40 new SB2, and 8 new SB3; see \cite{sota14}  for the latest results).

\section{Survey design\label{cafe}}

The CAF\'E-BEANS (CAF\'E -- Binary Evolution Andalusian Northern Survey) was designed to complement OWN in the Northern hemisphere and exploit the experience gained from both OWN and the IACOB survey \cite{simon10,simon14}. It uses the new CAF\'E (Calar Alto Fiber-fed \'Echelle) spectrograph (see \cite{aceituno}) on the 2.2~m telescope, at Calar Alto, Spain. The instrument is a single-fibre, high-resolution ($R\sim70\,000$) spectrograph covering the wavelength range 365\,--\,920 nm.

 The target list for  CAF\'E-BEANS was built in two steps. Initially, we considered all O-type stars included in the Galactic O Star Catalog (GOSC\footnote{{\tt http://ssg.iaa.es/en/content/galactic-o-star-catalog/}}) with $B<8$ and declination above $+20^{\circ}$ degrees. These two criteria guarantee respectively homogeneity and lack of repetition with the OWN survey. This first step yielded a total of 60 stars. When combined with OWN, this part
of the sample will provide a magnitude-limited sample for which we will
determine the multiplicity fraction. In addition, our sample will complement the results of the IACOB survey, which concentrates on determining parameters of apparently single O-type stars (see Sim\'on-D\'{\i}az et al. in this proceedings, and \cite{simon14}). 

In the second step, we added several O-type stars dimmer than $B=8$ that have already been observed to display radial velocity variations but have no orbital solutions (or poor quality ones). All the stars that displayed signs of binarity in IACOB spectra were included. We also added a few dim stars of special astrophysical interests (e.g.\ stars of suspected very high mass) that provide the {\it a priori} most interesting cases. This second step yielded 40 additional stars.

We aim at obtaining on average ten epochs for these 100 O-type stars. For objects that do not show evidence for variability, we will only obtain 3\,--\,4 epochs, while more than 12 will be taken for confirmed binaries, to solve their orbits. The survey was started in the autumn of 2012, and 5 nights have been allocated each semester since then, with very successful observations, except in F2013, when the survey was stopped for a few months by a major instrumental failure with CAF\'E, which was later compensated with extra time.

For a large number of systems, we had no {\it a priori} information on the orbital characteristics. Indeed, one of the main aims of the survey is characterising the distribution of orbital parameters, because there are no clear expectations from theoretical grounds. In general, objects with short orbits ($P<10$~d) will tend to be circularised \cite{north03}. When the orbit is circular, determination of orbital parameters is easier. As soon as a preliminary orbital solution can be obtained, we can concentrate on obtaining observations close to the quadratures. For longer periods, eccentricities may be rather high. The sample of systems with solutions is still too small to predict a distribution.

For orbital periods $P<15$~d, systems with similar components ($q\geq0.3$) appear as SB2 and give separations $\geq 100\:{\rm km}\,{\rm s}^{-1}$, except for very unfavourable configurations. Systems with extreme mass ratios, appear as SB1, but the cross-correlation technique can find variations of the order of 20\% of the velocity resolution (i.e.\ $\sim1\:{\rm km}\,{\rm s}^{-1}$), meaning that again we expect to miss only systems with the most unfavourable inclinations. 

For very long orbital periods, we are mostly sensible to systems with moderate $q$ (except for favourable inclinations, i.e.\ close to $90^{\circ}$).
The longest periods that are detectable with spectroscopic techniques are $\sim1-2$~years, corresponding to velocity separations of $\sim5\:{\rm km}\,{\rm s}^{-1}$ (i.e.\ the instrumental resolution) for systems with similar masses in circular orbits. Observing for at least 6 semesters, our project is sensitive to periods of the order of one year. Eccentric systems display higher separations close to periastron, but the chances to detect them are smaller. 

In all, the fraction of binary systems with periods below $\sim100$~d that cannot be detected due to unfavourable inclinations or extreme mass ratios is expected to be small thanks to the very high spectral resolution. But we will have to correct the results of our sample for these biasing effects. Their importance can be estimated by generating Monte-Carlo simulations of large populations with different original distributions.

\section{State of the survey}

Our observing strategy, designed in close communication with Calar Alto staff, intends to maximise results and minimise observing time. In a given night, the priority list is built using an algorithm that takes into account visibility and, when available, preliminary orbital solutions. With this strategy, we can observe each target at the orbital phase that will provide more information for the orbital solution. As we accumulate more datapoints, the orbital solutions can be improved and the algorithm becomes more efficient.
 
Each morning, the support astronomers send the status report to the PI and a link to the ftp site where the raw observations are located. By midday, the observations are already reduced. We can then review the spectra and check if the required S/N has been achieved. This allows us to re-elaborate our priority lists for the following night (if, for example, the sky conditions resulted in a high-priority target having less S/N than desired). 

Such a strategy implies that all radial velocity curves are being built in parallel. Its main disadvantage is that we have not sampled full individual curves until pretty recently, thus preventing early publication of results for any given target. The current statistics for the programme are given in Table~1.
A particularly interesting result is that, of the 78 stars with 3 or more epochs, only 9 show no sign of radial velocity variability. For 13 objects, the results are inconclusive, and the observed radial velocity variations might be due to causes other than binarity. All the other stars show clear signs of being spectroscopic binaries. This may be to some degree a result of a bias in selecting the objects (as described above), but the number of binaries in the magnitude-limited sample is also very high.

\begin{table}[ht] 
\caption{Statistics of observation for CAF\'{E}-BEANS (updated September 2014)} 
\center
\begin{minipage}{0.5\textwidth}
\center
\begin{tabular}{ll} 
\hline\hline 
Observations & Stars \\  
\noalign{\smallskip}
\hline
\noalign{\smallskip}
 $\geq10$ epochs & 16 stars\\
4 $-$ 10 epochs & 33 stars\\
3 epochs & 29 stars\\
 $\leq3$ & Rest\\
\noalign{\smallskip}
\hline
\end{tabular} 
\end{minipage}
\label{tab1} 
\end{table}

A good example of the advantages of the strategy followed is our contribution to solving the orbit of the inner pair of the massive $\sigma$~Ori Aa, Ab, B triple system. After first characterization of the orbital properties of the high eccentricity system $\sigma$~Ori~Aa,Ab \cite{sigori}, the team has continued gathering high-resolution spectroscopic data, using several telescope facilities. In particular, thanks to the flexibility of the observations for the CAF\'E-BEANS project, we could intensively sample the radial velocity curve of the system during the periastron passage in March 2014. Combining these observations with an extensive dataset spanning more than five years, the orbital period of the system has been determined with high precision (143.198 $\pm$ 0.005 d; \cite{simon15})

The high resolution of the spectra is also providing very interesting results, for example revealing the presence of third components in some known (or even moderately well-studied) SB systems.

However, it is important to take into account that, even though some individual orbital solutions can be of high scientific interest, our main scientific objectives rest on the statistical treatment of the whole dataset. Moreover, to reach our final science objectives, once the project is finished, we will combine it with OWN, and then correct the resulting distributions of orbital parameters for the biases introduced by systems that cannot be detected due to unfavourable inclinations or extreme mass ratios. As mentioned above, this will be done by resorting to Monte-Carlo simulations. 

%
%
% Do not delete the next line
\small  % Do not delete
%
%%% Comment the following line if you do not have acknowledgments.
\section*{Acknowledgments}   % Do not delete if you declare acknowledgments
%
%%% ACKNOWLEDGMENTS
%%% ACKNOWLEDGMENTS
We are very grateful to the staff of the Calar Alto Observatory, who continue providing excellent support in spite of increasing difficulties and decreasing resources. We thank the time allocation committee for guaranteed Spanish time for their continued support to this project.

This research is partially supported by the Spanish Ministerio de Econom\'{\i}a y Competitividad (Mineco) under 
grants AYA2012-39364-C02-01/02, AYA2010-21697-C05-04, AYA2013-40611-P, and the European Union. SSD is also funded by the Mineco under Severo Ochoa SEV-2011-0187, and by the Canary Islands Government under grant PID2010119. RHB acknowledges support from FONDECYT Project 1140076. This research has made use of the Simbad, Vizier 
and Aladin services developed at the Centre de Donn\'ees Astronomiques de Strasbourg, France.
%
% Do not delete the next few lines

%
\end{document}